\documentclass[sn-mathphys,Numbered]{sn-jnl}

\usepackage{graphicx}%
\usepackage{multirow}%
\usepackage{amsmath,amssymb,amsfonts}%
\usepackage{amsthm}%
\usepackage{mathrsfs}%
\usepackage[title]{appendix}%
\usepackage{xcolor}%
\usepackage{textcomp}%
\usepackage{manyfoot}%
\usepackage{booktabs}%
\usepackage{algorithm}%
\usepackage{algorithmicx}%
\usepackage{algpseudocode}%
\usepackage{listings}%
\usepackage{rotating}
\usepackage{array}
\usepackage{hhline, makecell,tabularx}
\usepackage{multirow}
\usepackage{breqn}
\usepackage{ae,aecompl}
\usepackage{subcaption}
\usepackage[T1]{fontenc}
\usepackage{longtable}
\usepackage{float}
\usepackage{mathtools}
\usepackage[italicdiff]{physics}

\theoremstyle{thmstyleone}%
\theoremstyle{thmstyletwo}%

\theoremstyle{thmstylethree}%

\begin{document}

\title[Article Title]{New Insight Concerning Primordial Lithium Production}


\author*[1]{\fnm{Tahani} \sur{Makki}}\email{trm03@mail.aub.edu}

\author[2]{\fnm{Mounib} \sur{El Eid}}\email{meid@aub.edu.lb}
\equalcont{These authors contributed equally to this work.}

\author[3]{\fnm{Grant} \sur{Mathews}}\email{gmathews@nd.edu}
\equalcont{These authors contributed equally to this work.}

\affil*[1]{\orgdiv{Department of Physics}, \orgname{American University of Beirut (Alumna)}, \orgaddress{\street{Bliss street}, \city{Beirut}, \country{Lebanon}}}

\affil[2]{\orgdiv{Department of Physics}, \orgname{American University of Beirut (emiritus)}, \orgaddress{\street{Bliss street}, \city{Beirut}, \country{Lebanon}}}

\affil[3]{\orgdiv{Department of Physics, Center for Astrophysics}, \orgname{University of Notre Dame}, \orgaddress{\street{Notre Dame}, \city{South Bend}, \postcode{46556}, \state{Indiana}, \country{USA}}}


\abstract{To constrain the universe before recombination (380000 years after the Big Bang), we mostly rely on the measurements of the primordial abundances that indicate the first insight into the thermal history of the universe. The first production of light elements is obtained by the Big Bang Nucleosynthesis (BBN). The production of the elements D, ${}^3$He, and ${}^4$He during BBN matches well the observations; however, the production of lithium (${}^7$Li) based on the Standard Big Bang Nucleosynthesis (SBBN) is found to be higher by about a factor of three than the observed abundance from metal-poor halo stars. This so-called "Cosmological Lithium Problem" is still elusive and needs to be resolved. One important attempt to resolve this problem is to invoke a non-standard description of the SBBN to decrease the lithium abundance.
In our previous work, we encountered a problem that the decrease in the ${}^7$Li abundance requires an increase in the deuterium abundance to maximum values that are not accepted by observations.
In the present work, a decrease in the lithium abundance could be achieved without maximizing the deuterium abundance by modifying the time-temperature relation in the range $(4.3-9.1)\times 10^{8}$ K during the nucleosynthesis process. This range is crucial to reducing the strong correlation between lithium and deuterium production. The main conclusion of the present work is that the ${}^7$Li abundance in the atmospheres of metal-poor stars cannot be analyzed without considering possible modifications to the primordial nucleosynthesis.}




\maketitle

\section{Introduction}\label{sec1}
The standard Big Bang nucleosynthesis (SBBN) is the production site of the light elements (D, ${}^3$He, ${}^4$He, ${}^7$Li, ${}^7$Be) before the formation of the first stars after the "dark age" which followed the recombination epoch, 380000 years after the Big Bang. The SBBN is a well-established theory since it depends on a single parameter the baryon-to-photon ratio ($\eta$) which has been well determined by the Planck satellite collaboration. The resulting abundances in the case of D and ${}^4$He match the observed abundances in the atmospheres of very metal-poor halo stars. However, the abundance of ${}^7$Li is found to be higher by at least a factor of three. This is termed as the "Cosmological Lithium Problem", which is not yet resolved.
Various works \cite{Goudelis 2016,Hou2017,Jedamzik2004a,Jedamzik2004b,Kusakabe2017,Mathews1990,Scherrer1,Scherrer2,Yamazaki2017} have attempted to resolve that problem, and the main result of these investigations was that the decrease in the lithium abundance was linked to an increase in the deuterium abundance to values not compatible with observations in stars. The drawback of that link is that deuterium is easily destroyed in stars at temperatures exceeding $10^6$ K. In our previous work \cite{Makki1}, we attempted to resolve the lithium problem by varying the number of neutrinos, their chemical potentials, and their temperature. In addition, the effect of dark fluid and photon cooling with axion dark matter \cite{Makki2} was included. However, this approach failed to relax the strong correlation between deuterium and lithium production.
In the present work, a decrease of the lithium abundance that matches closely the observations is possible without increasing the deuterium abundance to values contradicting the observations. As described in Sect. 4, this is obtained by modifying the time-temperature relation in the range $(4.3-9.1)\times10^8$ K when adding a dark entropy component. Sect. 2 introduces the SBBN. In Sect. 3, various suggestions are outlined to resolve the lithium problem. In Sect. 4 and 5, we have introduced our new approach to have a better understanding of the decrease of the lithium abundance resulting from the modified SBBN. Sect. 6 summarizes the conclusion.

\section{Standard Big Bang Nucleosynthesis (SBBN)}
\begin{figure}
    \centering
	\includegraphics[width=1\textwidth]{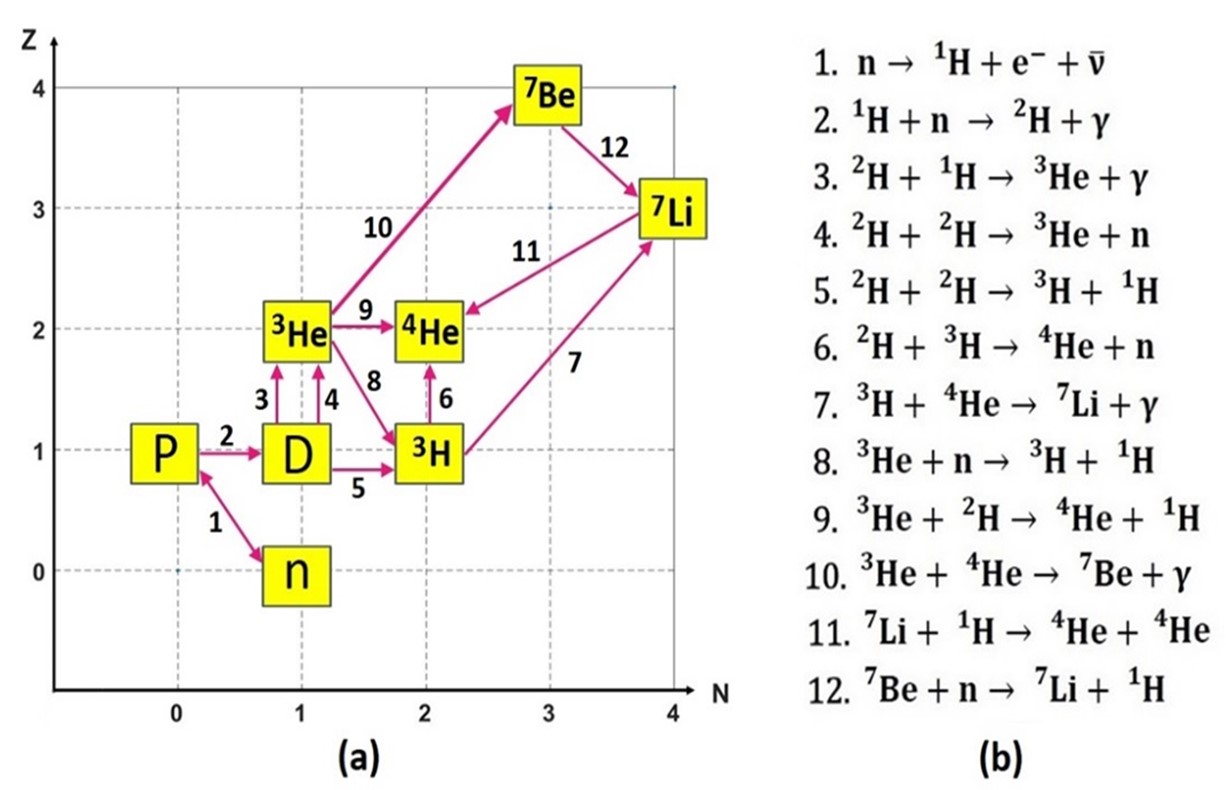}
    \caption{Network of the most important nuclear reactions included during primordial nucleosynthesis.}
    \label{Fig1}
\end{figure}

The SBBN model is described by the competition between the expansion time scale of the universe and the lifetime of the involved reaction rates. In particular, the weak interaction rates are sensitive to the expansion rate, so they will freeze out when their rates become less than the time scale of the universe.
Fig.(\ref{Fig1}) displays the network of the nuclear reactions, which we have used in the numerical simulations using a modified solver called "AlterBBN" \cite{AlterBBN}. This code requires the baryon-to-photon ratio $\eta$, the neutron lifetime $\tau_{n}$, and the relativistic effective degrees of freedom $N_{eff}$ as input. However, several modifications were required for the treatment of the non-standard Big Bang Nucleosynthesis (BBN). This concerns additional effective degrees of freedom ($N_{eff}$), neutrino degeneracy parameters $(\beta_{\nu_{e}}$,$\beta_{\nu_{\mu}}$,$\beta_{\nu_{\tau}})$, and dark components.
Table~\ref{Table1} includes the results obtained in this work using updated nuclear reaction rates, which lead to the lowest lithium production by the SBBN.
These calculations are in agreement with previous works \cite{Cuburt 2016, Pitrou 2018}. Table~\ref{Table1} shows that the ratios of the abundances normalized to hydrogen are in good agreement with observations, except for lithium which is overproduced by the SBBN.

\begin{table}[h]
\centering
\caption{The SBBN abundance ratios normalized to hydrogen obtained in the present work and by several other works as indicated.}
\begin{tabular}{@{}lllll@{}}\hline
 & This work & Pitrou et al. (2018) \cite{Pitrou 2018}& Cyburt et al. (2016) \cite{Cuburt 2016}  & Observations\\ \hline
Yp & 0.2461 $\pm$ 0.0002 & 0.24709 $\pm$ 0.00017 &0.24709 $\pm$ 0.00025& 0.2449 $\pm$ 0.0040 \cite{Aver 2015} \\ \hline
D/H $\times 10^{5}$ & 2.653 $\pm$ 0.123 & 2.459 $\pm$ 0.036 & 2.58 $\pm$ 0.13  & 2.58 $\pm$ 0.07 \cite{Cooke 2014}\\ \hline
${}^{3}He/H \times 10^{5}$ & 1.017 $\pm$ 0.053  & 1.074 $\pm$ 0.026 & 1.0039 $\pm$ 0.0090 & 1.1 $\pm$ 0.2 \cite{Bania 2002} \\ \hline
${}^{7}Li/H \times 10^{10}$ & 4.284 $\pm$ 0.378 & 5.623 $\pm$ 0.247 & 4.68 $\pm$ 0.67& $1.58^{+0.35}_{-0.28}$ \cite{Sbordone 2010}\\ \hline
\end{tabular}
\label{Table1}
\end{table}

\section{Different insights in the Cosmological Lithium Problem}
\begin{figure}
\centering
	\makebox[\textwidth]{\includegraphics[width=1.1\textwidth, height=8 cm]{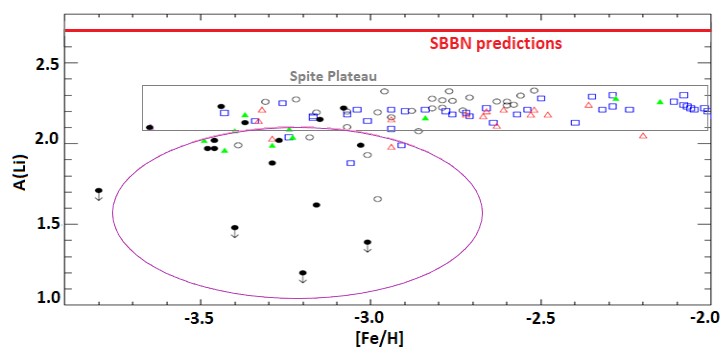}}
    \caption{Observed \cite{Li observations references, Fabio Iocco} and predicted abundances $A(Li)= log({}^{7}Li/H)+12$ as a function of metallicity characterized by  $[Fe/H] = log(N_{Fe}/N_{H})_{star} - log(N_{Fe}/N_{H})_{sun}$, where the N's are number densities. Note that $[Fe/H]=0$ is solar. Notice the difference between the predicted lithium abundance and the observed ones especially at very low metallicity.}
    \label{Fig2}
\end{figure}

An important observation indicating the discrepancy between the SBBN predictions and observations of the lithium abundance in the atmospheres of very metal-poor halo stars is called the  Spite Plateau, as shown in Fig.(\ref{Fig2}). This plateau exhibits single values with small dispersion in the metallicity range $-3.0\lesssim [Fe/H] \lesssim -2.0$ and the abundance range $2.1\lesssim A(Li)\lesssim 2.4$ (Fig.(\ref{Fig2}) explains these quantities). As Fig.(\ref{Fig2}) shows, the lithium abundance predicted by the SBBN is higher by a factor of at least three compared with the plateau level. In addition, that figure depicts a significant dispersion of the abundances below $[Fe/H]=-3.0$, a "melting" of the plateau behavior.
It is rather challenging to figure out one single reason to understand the behavior of the lithium abundances described above. Possible attempts are as follows.\\

\subsection{Stellar evolution aspects}
The main question is: what is the origin of lithium in the stellar atmospheres? One would not necessarily expect that element in the stellar atmosphere to indicate the original composition. The existence of the plateau led to the view that it could represent the primordial values \cite{Field 2011}. However, lithium is very fragile because it is destroyed at a temperature T $\sim 2.5\times 10^{6}$ K. Given that the stellar atmosphere is colder, convective mixing is needed to possibly bring the lithium down to hotter layers. But how deep could this penetration be, called overshooting? This mechanism is a non-local convective mixing beyond the border of the convectively unstable region according to the "Schwarzschild criterion". The overshooting is commonly a parametrized process \cite{Bressan 2015, Fu 2015}. The authors in Ref.\cite{Bressan 2015} assumed that the SBBN is correct, and they used overshooting, microscopic diffusion, and residual mass accretion during the pre-main sequence evolution of very metal-poor stars of masses 0.57-0.60 M$_\odot$. Their approach led to obtaining the spite plateau, but it does not explain the dispersion below it. According to Ref.\cite{Brian2022}, they also suggested an astrophysical solution without modifying the SBBN. Such a solution needs to be supported by future observations and should consider the plateau level and the dispersion below it. Then, it is worth to include the non-standard effects we describe in the present work.

\subsection{Nuclear physics aspects}
The observed lithium abundances in metal-poor halo stars depend on the nuclear reaction rates involved in the SBBN, especially the reactions (10) and (12) listed in Fig.(\ref{Fig1}). Reaction (10) produces ${}^{7}$Be, which enables the production of ${}^{7}$Li via electron capture after the end of BBN. In addition, the reactions responsible for the production and destruction of deuterium can indirectly influence the final abundance of ${}^{7}$Li since these two elements are directly related as shown in Ref.\cite{Mukhanov2008} and our recent calculations \cite{Makki1}.
It is clear from the network shown in Fig.(\ref{Fig1}) that the three fundamental processes (strong nuclear, weak nuclear, and electromagnetic) indicate how fundamental the Big Bang Nucleosynthesis (BBN) is.

\subsection{Modification of the SBBN}
A third attempt is to investigate a modification of the SBBN. Various attempts have been proposed to resolve the lithium problem dealing with inhomogeneous nucleosynthesis \cite{Mathews1990, Nakamura 2017}, the decay of massive particles during the SBBN \cite{Goudelis 2016, Jedamzik2004a, Jedamzik2004b}, and modification of the Maxwell-Boltzmann distribution of nuclei during BBN \cite{Hou2017}. Another approach to reducing the primordial lithium abundance was to assume photon cooling, the decay of long-lived X particles, and fluctuations of a primordial magnetic field \cite{Yamazaki2017}.
In our previous work \cite{Makki1, Makki2},  we considered a non-standard treatment of the SBBN to find out how far we can reduce the lithium abundance. We have included non-standard neutrino properties and dark components.  We have been able to reduce the lithium abundance but at the expense of increasing the deuterium abundance, which was not compatible with observation. In the present work, we describe a reasonable way to avoid this drawback.

\section{Modification of the entropy during BBN}
\subsection{General comments}
Dark matter is a crucial component in the universe whose constituents have been the focus of many works for a long time. Different candidates of dark matter are proposed, such as gravitino dark matter \cite{Daniel2006}, sterile neutrinos \cite{Kevork2001}, axions \cite{Duffy2009}, and unifying dark matter with dark energy \cite{Arbey2010}. Such particles may decay into other particles during BBN. For example, the decay of cold dark matter was examined to alleviate the tension between the actual Hubble constant (H$_{0}$) measured by cosmic microwave radiation and the one obtained from the distance ladder measurements from SNIa \cite{Vattis2019}.
Gravitino decay modes produce hadronic and electromagnetic spectra, which interact with the background nuclei during BBN and affect their abundance \cite{Feng2012}. However, that decay is constrained by the observed abundances of D and ${}^{4}$He; then, for large gravitino mass (equal to or greater than 3 Tev) and a lifetime of around 103 seconds, the lithium problem is alleviated. It is beyond the scope of the present work to discuss all possibilities of modifying the SBBN (see Ref.\cite{Justin2012} for details). In any case, possible suggestions have to produce the observed abundances of D and ${}^{4}$He. In the next paragraph, we focus on investigating the impact of dark entropy on the resulting lithium abundance during BBN. We note that the adopted description of dark matter and entropy in the present work relies on the conventional model of dark matter; this is to illuminate its effect on the primordial lithium production. We think that more investigations are needed in that respect.
\subsection{Variation of the entropy during BBN}
As we will see, it is important to vary the entropy content during BBN since this affects the energy conservation equation and consequently the time-temperature relation. Such a variation could result from the decay or interaction of some weakly interacting particles with the background radiation or any other source that could be modelled as a dark component. With the  adiabatic expansion of the universe and no
modification of its entropy content, the energy conservation equation reads:
\begin{equation}\label{Eq.l}
\dv{t}(\rho_{tot}\, a^{3}) + P_{tot} \dv{t}(a^{3})= 0,
\end{equation}
where "a" is the scale parameter, $\rho_{tot}$ and $P_{tot}$ are the total energy density and the corresponding pressure of all constituents, namely, photons, neutrinos, baryons, electrons, and positrons.
In BBN codes, Eq.(\ref{Eq.l}) is transformed into the following relation:
\begin{equation}\label{Eq.2}
\dv{T}(\ln a^{3}) = -\frac{\dv{\rho_{tot}}{T}}{\rho_{tot} + P_{tot}},
\end{equation}
Having $\dv{T}(\ln a^{3})$, the time-temperature relation is implemented in BBN codes as follows:
\begin{equation}\label{Eq.3}
\dv{T}{t}=3\,H\Big/\dv{T}(\ln a^{3}),
\end{equation}
where H is the Hubble parameter. Introducing the dark entropy into Eq.(\ref{Eq.l}), it takes the following form:
\begin{equation}\label{Eq.4}
\dv{t}(\rho_{tot}\, a^{3}) + P_{tot}\dv{t}(a^{3}) + T \dv{t}(f_{D}\, a^{3}) = 0
\end{equation}
In this equation, we represent the modification of the energy conservation equation by a function $f_{D}$ that could take many forms. In this work, $f_{D}$ is the temperature-dependent dark entropy density. This function is used to modify Eq.(\ref{Eq.2}) which becomes:
\begin{equation}\label{Eq.5}
\dv{T}(\ln a^{3}) = -\frac{\dv{\rho_{tot}}{T} + T \dv{f_{D}}{T}}{\rho_{tot} + P_{tot} + T f_{D}}.
\end{equation}

Varying the entropy content of the early universe was studied in our previous work \cite{Makki2}. However, in the present work, the variation of the entropy is restricted to the temperature range $T =(4.3-9.1)\times 10^{8}$ K. This modification will affect the final element abundances. We adopt the proposal in Ref.\cite{Arbey and Mahmoudi 2010}, where a unified fluid is adopted to describe dark energy and dark matter as two different aspects of the same component. This is represented by temperature-dependent components of dark energy density $\rho_{D}$ and dark entropy $s_{D}$ as follows:
\begin{equation}\label{Eq.6}
\rho_{D}(T) = k_{\rho}\times \rho_{rad}(T_{0})\times \left(\frac{T}{T_{0}}\right)^{n_{\rho}}
\end{equation}

\begin{equation}\label{Eq.7}
s_{D}(T) = k_{s}\times s_{rad}(T_{0})\times \left(\frac{T}{T_{0}}\right)^{n_s}
\end{equation}
where $\rho_{rad}(T_{0})$ and $s_{rad}(T_{0})$ are the radiation and entropy energy densities respectively at $T_{0} = 1.0$ Mev = 11.6 GK, $k_{\rho}$ is the ratio of the effective dark fluid energy density over the total radiation at T$_{0}$, and $k_{s}$ is the ratio of the effective dark fluid entropy density over the total entropy density at T$_{0}$. The exponents $n_{\rho}$ and $n_{s}$ characterize the behaviors of dark energy density and entropy; for example, $n_{s}=3$ corresponds to a radiation behavior, while $n_{s}=1$  characterizes the entropy behavior that appears in reheating models. In this case, $f_{D} = s_{D}$ in Eq.(\ref{Eq.4}).
The effects of adding a dark entropy component in the temperature range $(4.3-9.1)\times10^8$ K and from the beginning to the end of BBN are summarized in Table~\ref{Table2}.

\begin{table}[h]
\centering
\caption{Effect of dark entropy on light elements when inserted in the temperature range $T=(4.3-9.1)\times 10^{8}$ K and from the beginning till the end of BBN ($T=0.001-100$ GK). We show the parameters of dark entropy as implemented in AlterBBN code and the resulted abundances of light elements.}
\begin{tabular}{@{}lllll@{}}\hline
Dark entropy parameters & Temperature range &Yp & D/H $\times 10^{5}$ &${}^{7}Li/H\times 10^{10}$\\[2pt] \hline
$n_{s} = 1$, $k_{s} = 10^{9}$& $T=(4.3-9.1)\times 10^{8}$ K & $0.2516 \pm 0.0005$ & $2.690 \pm 0.202$ & $2.249 \pm 0.374$ \\ $n_{s} = 1$, $k_{s} = 10^{9}$&$T=0.001-100$ GK & $0.2993 \pm 0.0001$ & $1.517 \pm 0.081$ & $10.06 \pm 0.925$ \\\hline
$ n_s = 0.8,  k_s = 9\times 10^{8}$ & $T =(4.3-9.1)\times 10^{8}$ K & $0.2506 \pm 0.0004$ & $2.725\pm0.216$ & $2.285 \pm 0.434$ \\$ n_s = 0.8,  k_s = 9\times 10^{8}$  &  $T=0.001-100$ GK & $0.2791\pm0.0001$ & $2.360\pm0.112$ & $5.725\pm0.515$ \\\hline
$ n_s = 0.5, k_s = 10^{10}$ & $T =(4.3-9.1)\times 10^{8}$ K & $0.2531 \pm 0.0006$ & $3.091\pm0.312$ & $1.544\pm0.374$ \\
$ n_s = 0.5, k_s = 10^{10}$ & $T=0.001-100$ GK & $0.3001 \pm 0.0001$ & $2.974\pm0.126$ & $4.734\pm0.420$\\\hline
\multicolumn{2}{l}{Standard Big Bang Nucleosynthesis (SBBN)}& $0.2461\pm0.0002$ & $2.653 \pm$ 0.123& $4.284 \pm 0.378$\\ \hline
\end{tabular}
\label{Table2}
\end{table}

The results in Table~\ref{Table2} indicate that using the description of the dark entropy as in Eq.(\ref{Eq.7}), the decrease of the lithium abundance while keeping the helium and deuterium abundances compatible with the observations ($ n_s = 1,  k_s = 9\times 10^{9}$) is possible only if we apply Eq.(\ref{Eq.7}) in the temperature range $T=(4.3-9.1)\times 10^{8}$ K. The consequence of this restriction alters the time-temperature relation and affects the nuclear reaction rates shown in Fig.(\ref{Fig1}). Other combinations of $n_s$ and $k_s$ are possible to achieve similar results ($ n_s = 0.8,  k_s = 9\times 10^{8}$). However, if we want to reach the lower level of the plateau, or even go below it, we must relax the constraint on deuterium. In this case, we can argue that this increase in
deuterium could be accepted because deuterium is fragile and can be destroyed in stars.
For the same set of parameters, adding the dark entropy from the beginning till the end of BBN ($T = 0.001 - 100$ GK) violates the observational constraints on D and ${}^{4}$He.
We emphasize that the resulting lithium abundance is the sum of ${}^{7}$Li and ${}^{7}$Be; more precisely, ${}^{7}$Be has a crucial role in determining the lithium abundance since it represents more than $90\%$ of the final lithium abundance. In the following, we present the abundances of light elements as a function of temperature resulting from the present treatment.
\subsection{Effect of dark entropy on the ${}^{4}$He abundance}
\begin{figure}[H]
	\includegraphics[width=1\textwidth,height=7 cm]{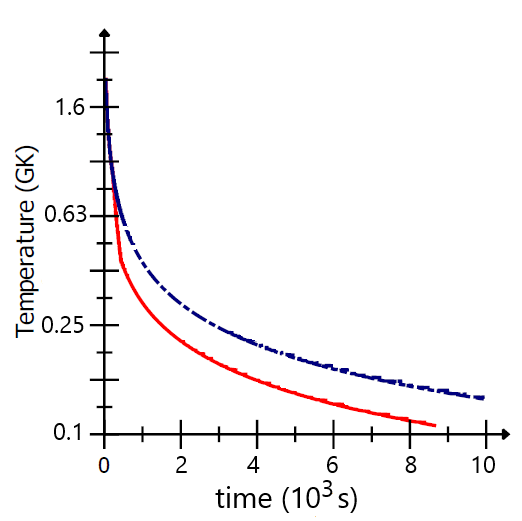}
    \caption{The variation of the temperature as a function of time during the SBBN (blue dashed line) and when including the dark entropy component in the temperature range $T = (4.3-9.1)\times 10^{8}$ K (red solid line).}
    \label{Fig2'}
  \end{figure}

\begin{figure}[H]
\makebox[\textwidth]{\includegraphics[width=1.2\textwidth, height = 7.5cm]{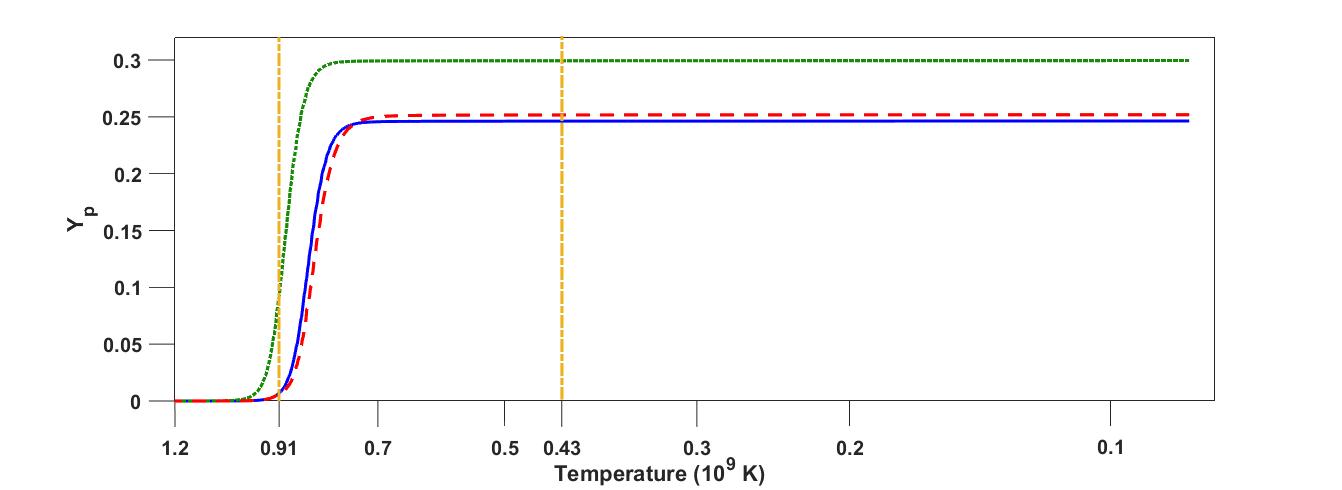}}
    \caption{The abundance of helium-4 based on the SBBN (blue solid line), when including the dark entropy during the BBN epoch (green dotted line), and when including the dark entropy in the temperature range $T = (4.3-9.1)\times 10^{8}$ K (red dashed line).}
    \label{Fig3}
  \end{figure}

Fig.(\ref{Fig3}) shows the helium abundance as a function of temperature. Inserting the dark entropy component during BBN does not alter the shapes of the abundance profile. Including this entropy component during the whole BBN epoch ($T = 0.001 - 100$ GK) increases the helium abundance, which contradicts the observations. However, the helium abundance agrees with the value from the SBBN and observations if the dark entropy is applied in the range $T = (4.3-9.1)\times 10^{8}$ K. To understand this result, we note the following.
Helium is affected by two important stages:
(i) freeze-out of weak interaction reactions ($n +\nu_{e}\leftrightarrow p + e^{-}$, $n +e^{+}\leftrightarrow p + \bar{\nu}_{e}$, $n \leftrightarrow p + e^{-} + \bar{\nu}_{e}$), (ii) deuterium bottleneck (its maximum) during BBN. Since the dark entropy causes a faster decrease of the temperature (see Fig.~\ref{Fig2'}) during BBN, the deuterium abundance reaches its bottleneck at a higher temperature or earlier (see green dotted curve in Fig.(\ref{Fig4})), which leads to an increase in the helium abundance (green dotted curve in Fig.(\ref{Fig3})). This relation between helium and deuterium bottleneck is given explicitly in Ref.\cite{Mukhanov2008} where the final abundance of helium is proportional to $exp(-t_{N}/\tau_{n})$, with $t_{N}$ is the time of the bottleneck and $\tau_{n}$ is the neutron lifetime. However, restricting the dark entropy in the adopted range does not affect the helium abundance (red dashed line curve in Fig.(\ref{Fig3})). This is because the time of the deuterium bottleneck is not shifted although the abundance of the bottleneck is higher than that of the SBBN (red dashed line in Fig.(\ref{Fig4})). In addition, the freeze-out temperature is not affected because it happened at higher temperatures, so the ${}^{4}$He abundance remains in the acceptable range as Fig.(\ref{Fig3}) shows.

\subsection{Effect of dark entropy on the deuterium abundance}
\begin{figure}[H]
 \centering
\makebox[\textwidth]{\includegraphics[width=1.2\textwidth, height=7.5cm]{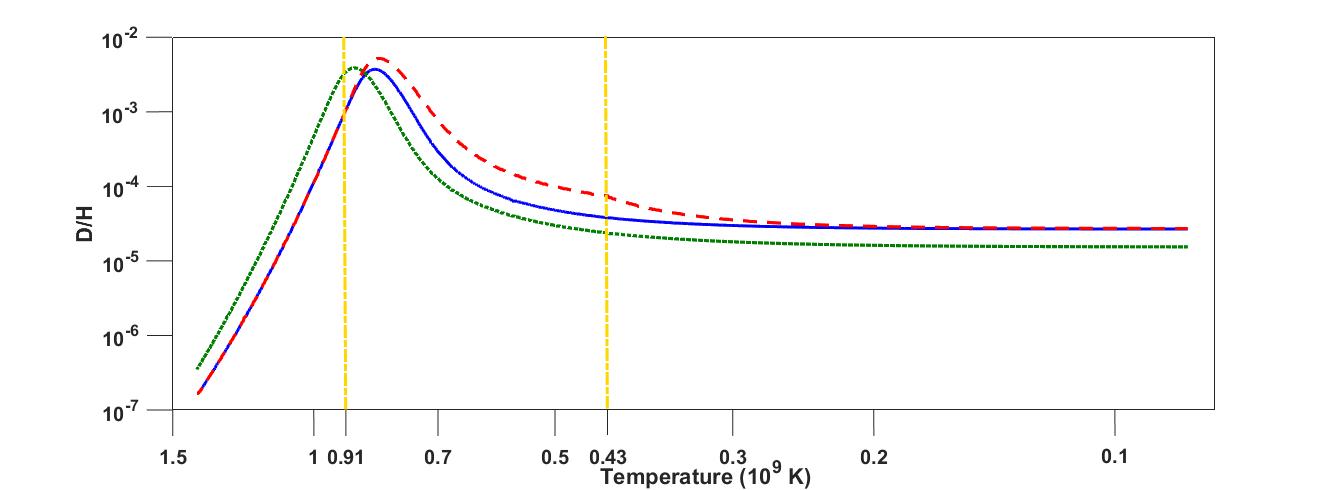}}
 \caption{The deuterium abundance based on the SBBN (blue solid line), when including the dark entropy during the BBN epoch (green dotted line), and when including the dark entropy in the temperature range $T = (4.3-9.1)\times 10^{8}$ K (red dashed line).}
 \label{Fig4}
\end{figure}

\begin{figure}[H]
\centering
\makebox[\textwidth]{\includegraphics[width=1.1\textwidth,height=7cm]{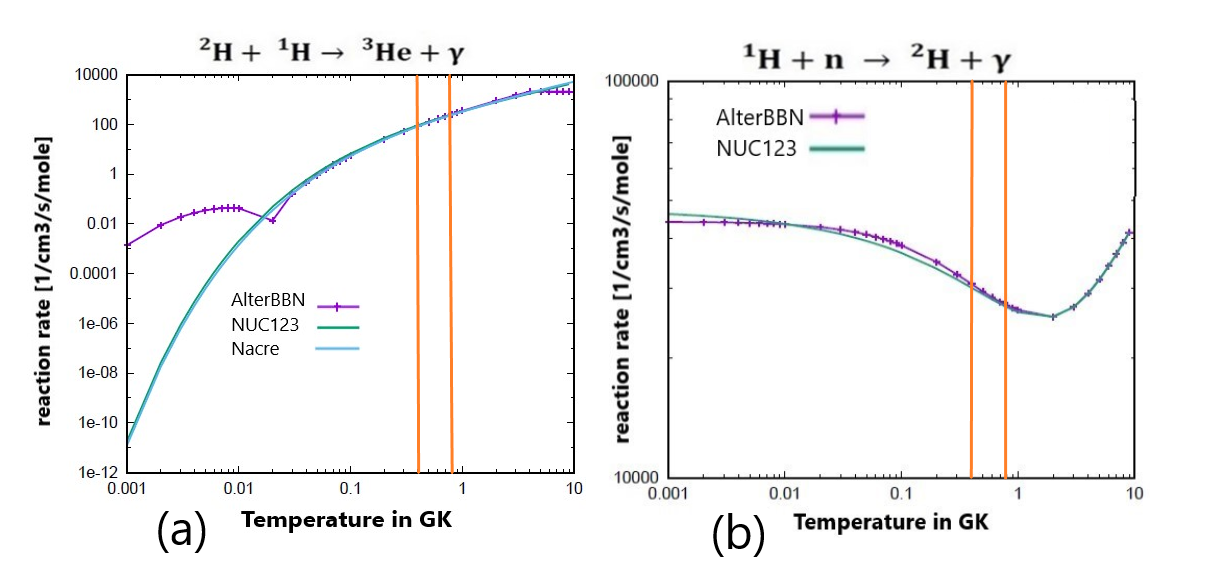}}
\caption{Example of two reaction rates responsible for the destruction (a) and production (b) of deuterium. The vertical lines represent the restricted temperature range.}
\label{Fig5}
\end{figure}

It is emphasized that  the modification of the entropy content took place before two important phases:
(i) the deuterium bottleneck (its maximum), and (ii) when the neutron abundance becomes comparable to that of deuterium.
 As seen in Fig.(\ref{Fig4}), the deuterium abundance increases at the bottleneck but converges to its standard value below 0.43 GK (the red dashed curve converges to the blue solid curve). It is clearly shown in Table~\ref{Table2} that the final deuterium abundance $2.690\times10^{-5}$ is approximately equal to the resulting abundance of the SBBN within the uncertainties. This result is due to the efficiency of the reaction rates shown in Fig.(\ref{Fig5}).

\subsection{Effect of dark entropy on ${}^{7}$Be and ${}^{7}$Li abundances}
\begin{figure}
\centering
\makebox[\textwidth]{\includegraphics[width=1.2\textwidth, height=7.5 cm]{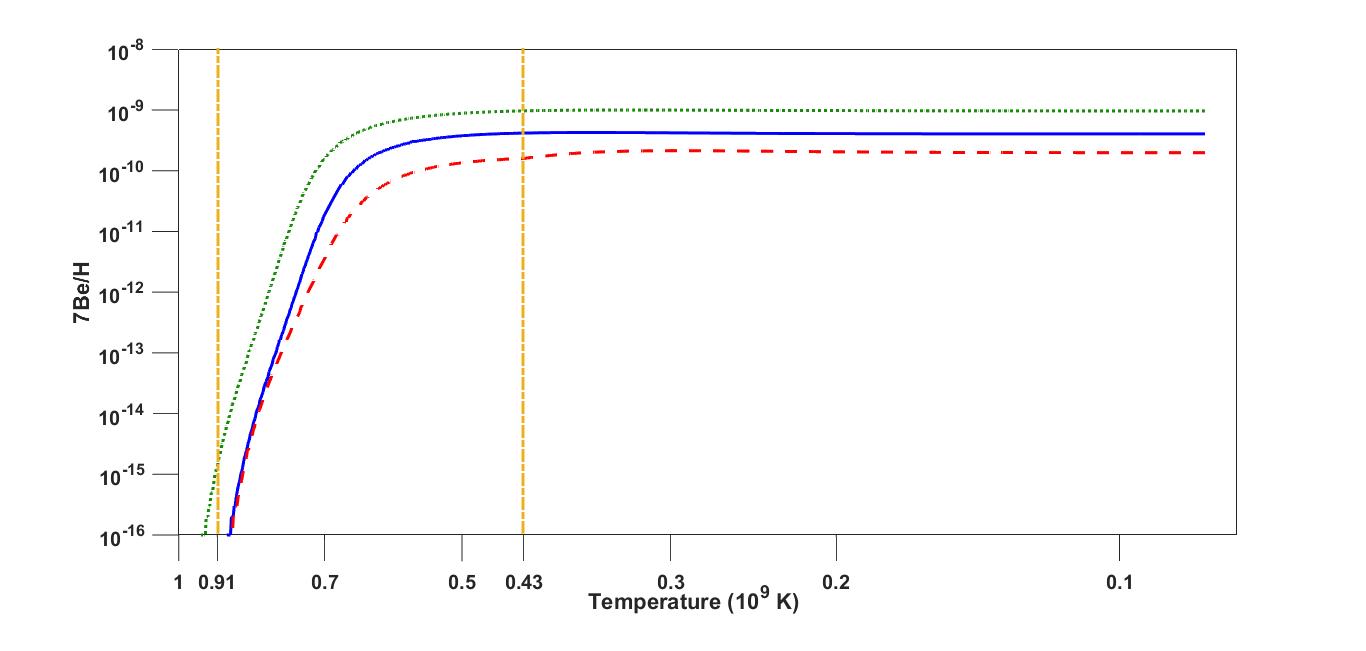}}
    \caption{The abundance of beryllium based on the SBBN (blue solid line), when including the dark entropy during the BBN epoch (green dotted line), and when including the dark entropy in the temperature range $T = (4.3-9.1)\times 10^{8}$ K (red dashed line).}
    \label{Fig6}
\end{figure}

\begin{figure}[H]
  \centering
 \makebox[\textwidth]{\subfloat[]{\label{figure7a}\includegraphics[width=67mm, height=70mm]{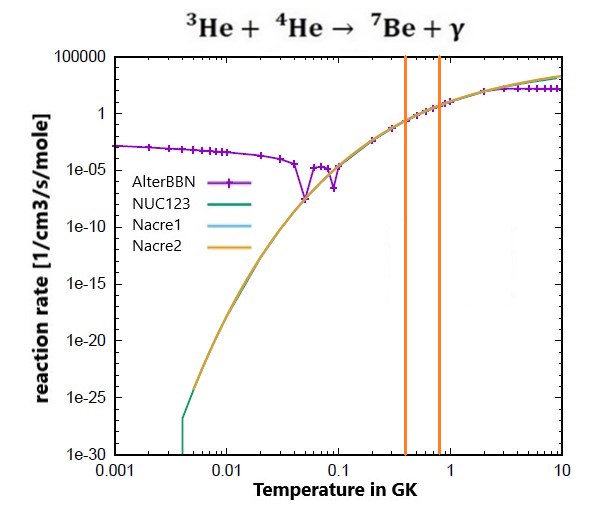}}
    \subfloat[]{\label{figure7b}\includegraphics[width=67mm, height=70mm]{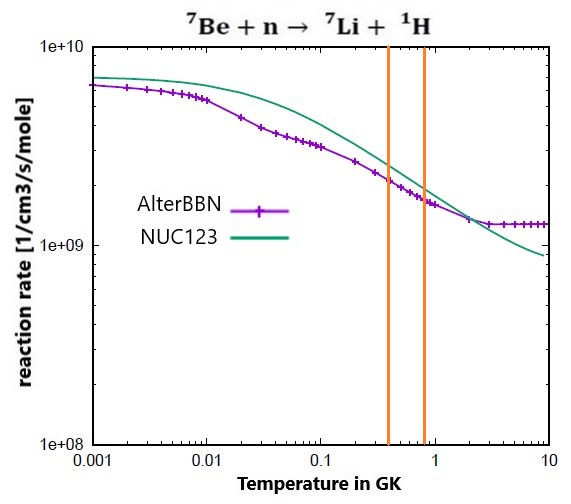}}}
  \caption{The main reaction rates responsible for the production (a) and destruction (b) of ${}^{7}$Be. The vertical lines represent the restricted temperature range.}
  \label{Figure7}
  \end{figure}

The lithium abundance decreases significantly to $2.249\times10^{-10}$ (see Table~\ref{Table2}) to match closely the observed plateau level. This is due to adding the dark entropy in the restricted temperature range $T = (4.3-9.1)\times 10^{8}$ K, which leads to a faster decrease in the temperature shown in Fig.~\ref{Fig2'}. Fig~\ref{Fig6} shows the abundance of ${}^{7}$Be as a function of temperature. A similar profile is obtained in all three cases: (a) SBBN, (b) dark entropy applied in the range of BBN, and (c) in the restricted temperature range we have adopted. In case (c), the resulting lower abundance reflects the behaviour of the reaction rates as shown in Fig~\ref{Figure7}. The inspection of Fig~\ref{figure7a} reveals that the production of ${}^{7}$Be through the reaction ${}^{3}He + {}^{4}He \longrightarrow {}^{7}Be + \gamma $ is not enough compared to the destruction by the reaction ${}^{7}Be + n \longrightarrow {}^{7}Li + p $ (Fig.~\ref{figure7b}).
The present investigation indicates the importance of the time evolution of the temperature and its impact on the reaction rates in determining the abundances of ${}^{7}$Be and consequently ${}^{7}$Li. Then, the desired decrease in the abundance of ${}^{7}$Li is achieved without affecting the observational constraints on helium and deuterium.

\section{Another view of the modification of the time-temperature relation}
In section 4, we modified the energy conservation equation given in Eq.~\ref{Eq.l}, which
affected the time-temperature relation. Here, we will directly perturb the time-temperature relation to emphasize that altering the energy conservation equation by a function $f_{D}$ implies a modification of the time-temperature relation. In this way, we could reach a substantial decrease in the lithium abundance without increasing the deuterium abundance using different parametrizations.
The perturbation is introduced  by a function $g(T)$ added to Eq.~\ref{Eq.3} in the range $T = (4.3-9.1)\times 10^{8}$ K, so that Eq.~\ref{Eq.3} becomes:
\begin{equation}\label{Eq.8}
\dv{T}{t} = 3\,H\Big/\dv{T}(\ln a^{3}) + 3\,H \times g(T),
\end{equation}
Interesting to find an expression to this function, for example by a linear fitting. Our trial yields:

\begin{equation}\label{Eq.9}
 g(T) = 1.64 -3.72\,T
\end{equation}
Modifying the time-temperature relation in this way leads to the results in Table~\ref{Table 3} showing that the lithium abundance has significantly decreased without altering the abundances of helium and deuterium from the SBBN.
 \begin{table}[h]
\centering
\caption {The effect of modifying the time-temperature relation of the early universe by a function g(T) in the temperature range $T = (4.3-9.1)\times 10^{8}$ K.}\label{Table 3}
\begin{tabular}{@{}lll@{}} \hline
 & SBBN & using g(T)\\ \hline
Yp & 0.2461$\pm$0.0002 & 0.2549$\pm 0.0008$ \\ \hline
D/H $\times 10^{5}$ & 2.653 $\pm$ 0.123 & 2.624 $\pm$ 0.119 \\ \hline
${}^{7}Li/H \times 10^{10}$ & 4.284 $\pm$ 0.378 & 2.204 $\pm$ 0.222\\ \hline
\end{tabular}
\label{Table3}
\end{table}
\section{Summary and Conclusion}
The standard Big Bang nucleosynthesis (SBBN) is a fundamental process in cosmology responsible for the formation of light elements up to lithium (${}^{7}$Li), which needs extended work to resolve the discrepancy between its predicted lithium abundance and the observed abundance in very metal-poor halo stars. To resolve this so-called "Cosmological Lithium Problem", we need an extension of the SBBN before drawing conclusions based on stellar modelling. In our previous works \cite{Makki1,Makki2}, we have implemented non-standard treatments of BBN by varying neutrino properties and adding dark components. We have achieved a decrease in the lithium abundance but only combined with an increase of the deuterium abundance, which does not match the accurate observations.
In the present work, we have invoked an additional treatment based on modifying the energy conservation equation by adding a dark entropy term (see Sect. 4.2). This resulted (Sect. 4 for details) in relaxing the tight connection between deuterium and lithium. With this approach, we figured out that the dark entropy term with its used parametrized form (Sect. 4.2) cannot be applied through the whole temperature range of the SBBN, but in a restricted range $T=(4.3-9.1)\times 10^{8}$ K. This restriction allows to reduce the lithium abundance to match closely the observations and satisfy the constraints on helium and deuterium (see Table~\ref{Table2}). Additionally, as seen in Sect. 5, we find a perturbation term g(T) describing the behavior of the time-temperature relation in the restricted range above.
Finally, we emphasize that including the effect of the dark entropy component during BBN, which affects the time-temperature relation and relevant nuclear reactions, is a crucial issue in resolving the cosmological lithium problem. Stellar evolution modelling cannot simply assume that the SBBN is just correct.

\bibliography{sn-bibliography}

\end{document}